\documentclass{article}
\usepackage[utf8]{inputenc}
\usepackage{graphicx,psfrag,epsfig,subfig,comment,array}
\usepackage{amsmath,amssymb,mathtools}
\usepackage{latexsym}
\usepackage{multicol,xspace}
\usepackage{float}
\usepackage{booktabs}
\usepackage{multirow}
\usepackage{rotating}
\usepackage{bm}
\usepackage{tabularx}
\usepackage{amsthm}
\usepackage{flushend}
\usepackage{algorithm,algpseudocode}
\usepackage{pbox,footnote}
\usepackage{slashbox,makecell}
\usepackage{diagbox}
\usepackage[english]{babel}
\usepackage{cite}
\usepackage{color}
\usepackage{footnote}

\def\Keywords{robust filtering, Huber's cost, correlated measurement, outliers.}

\newtheorem{remark}{Remark}

\gdef\thefootnote{\arabic{footnote}}\gdef\@@savethanks{}%
\def\sthanks#1{\gdef\thefootnote{\fnsymbol{footnote}}\@@savethanks{#1}}

\begin{document}

\title{Outlier-robust Kalman Filter in the Presence of Correlated Measurements}

\author{Hongwei~Wang,
        Yuanyuan~Liu,
        Wei~Zhang,
        and~Junyi Zuo% <-this % stops a space
\thanks{The first two authors contributed equally to this work.}
%\thanks{This work was supported by the China Scholarship Council and the National Natural Science Foundation of China (NSFC) under Grant 61473227 and 11472222.}
\thanks{Hongwei~Wang is with the National Key Laboratory of Science and Technology on Communications, University of Electronic Science and Technology of China, Chengdu 611731, China (email: tianhangxinxiang@163.com).}
\thanks{Yuanyuan~Liu, Wei Zhang and Junyi Zuo are with the School of Aeronautics, Northwestern Polytechnical University, Xi'an 710072, China.}
}% <-this % stops a space

\date{October 2020}

\maketitle

\begin{abstract}
  We consider the robust filtering problem for a state-space model with outliers in correlated measurements. We propose a new robust filtering framework to further improve the robustness of conventional robust filters. Specifically, the measurement fitting error is processed separately during the reweighting procedure, which differs from existing solutions where a jointly processed scheme is involved. Simulation results reveal that, under the same setup, the proposed method outperforms the existing robust filter when the outlier-contaminated measurements are correlated, while it has the same performance as the existing one in the presence of uncorrelated measurements since these two types of robust filters are equivalent under such a circumstance.
\end{abstract}
~\\

\noindent{\bf Keywords: \Keywords} 

\begin{comment}
\begin{keywords}
robust filtering, Huber's cost, correlated measurement, outliers.
\end{keywords}
\end{comment}

\section{Introduction}

State estimation for discrete-time stochastic dynamic systems using noisy measurements has been attracting significant attentions. It is frequently encountered in a variety of industrial appliances such as target tracking, information fusion, attitude determination, and many others. The celebrated Kalman filter (KF)~\cite{kalman1960new} and its nonlinear suboptimal extensions, e.g., the cubature Kalman filter (CKF)~\cite{arasaratnam2009cubature}, offer some feasible estimation for state-space models with Gaussian process and measurement noises. Nevertheless, these well-established techniques may potentially degrade for scenarios where measurements are contaminated by outliers. The main reason is that the aforementioned filtering algorithms, in essence, are derived from the linear minimum mean square error criterion which is sensitive to non-Gaussian noises owing to outliers~\cite{schick1994robust}. Therefore, considerable amounts of efforts have been devoted to enhancing the robustness of the filtering methods.

Multiple-model or Gaussian-mixture methods based techniques~\cite{faubel2009split,terejanu2011adaptive,gao2017interacting} and sequential Monte Carlo sampling methods~\cite{arulampalam2002tutorial} can in principle be employed to handle non-Gaussian noises. Nevertheless, the heavy computational burden makes them infeasible for real-time implementation. Another line of research is to utilized some heavy-tailed distributions, e.g., the Student's t distribution~\cite{agamennoni2012approximate} and Laplace distribution~\cite{wang2017laplace}, to model noises of outlier-contaminated measurements. The good performance of these methods is reached only when the parameters associated with the heavy-tailed distributions are properly selected, while in general it is lacking in the guideline for parameter selection.

%Other approaches for robust filtering, e.g., the H$_\infty$ filter and the outlier detect-and-reject filter, are also studied.

The M-estimation from robust statistics is also a common strategy for robust filtering. In~\cite{masreliez1977robust} the Huber cost was first introduced to a linear dynamic system to derive a robust KF by recasting the filtering problem as linear regression. Thereafter, this design methodology was extended to nonlinear systems via introducing a linearization procedure, resulting in several nonlinear robust Kalman filters, e.g.,~\cite{karlgaard2007huber,chang2012huber,chang2013robust}. Most recently, the nonlinear filtering problem was interpreted as nonlinear regression, and a nonlinear regression based robust filter was proposed~\cite{karlgaard2015nonlinear}. Furthermore, considering the fact that there are other robust cost functions besides Huber's cost, a unified form for robust filters based on M-estimation was presented in~\cite{chang2017unified}. The M-estimation based solutions, in essence, are to reweight the measurement noise covariance based on the normalized measurement fitting error (NMFE). Each component of the NMSE will affect each other when measurements are correlated. The value of normal components in the NMSE may be influenced by outlier-contaminated ones, resulting in an improper reweighting of the noise covariance and furthermore an information loss about the state.

In the present paper, we focus on developing a general robust filtering framework for dynamic systems with correlated measurements that may be disturbed by outliers. Specifically, we consider the measurement fitting error (MFE) rather than the NMSE. The MFE is processed separately during the reweighting procedure to avoid the negative influence that the outlier-contaminated elements have enforced on the normal ones, which differs from existing robust filters where the MFE is normalized and jointly processed. Numerical results show that, in the presence of outliers, the proposed scheme outperforms the existing robust filter with a same setup when measurements are correlated, while has the same performance when uncorrelated measurements are involved.

\section{Problem Formulation}

Consider the following nonlinear discrete dynamic system
\begin{align}
\bm x_t &= f(\bm x_{t-1}) + \bm v_t\notag\\
\bm y_{t} &= h(\bm x_t) +\bm w_{t}
\end{align}
where $\bm x_t\in \mathbb{R}^n$ is the state of interest at time instant $t$; $\bm y_t\in\mathbb{R}^m$ is the corresponding observation; $f(\cdot)$ and $h(\cdot)$ are some known mappings referred to the state evolution and observation procedure respectively; $\bm v_t$ is the process noise which is assumed to be a Gaussian $\mathcal{N}(0,\bm Q_t)$; and $\bm w_t$ denotes the measurement noise which follows a Gaussian distribution $\mathcal{N}(0,\bm R_t)$ in most cases. The initial state $\bm x_0$ is known a priori, i.e., $\bm x_0\sim\mathcal{N}(\hat{\bm x}_{0|0},\bm P_{0|0})$. In this work $\bm x_0$, $\bm v_t$, and $\bm w_t $ are assumed to be mutually independent.

Generally, the measurement $\bm y_t$ comes from either the multiple outputs of a specific sensor or the single output of several sensors. Therefore, $\bm y_t$ contains several components, which can be expressed as
\begin{align}
\bm y_t =
\begin{pmatrix}
y_{t,1}\\
\vdots\\
y_{t,m}
\end{pmatrix}
=
\begin{pmatrix}
h_1(\bm x_t)\\
\vdots\\
h_m(\bm x_t)
\end{pmatrix}
+
\begin{pmatrix}
w_{t,1}\\
\vdots\\
w_{t,m}
\end{pmatrix}
\end{align}
The elements in $\bm y_t$ might be correlated conditioned on the state $\bm x_t$, resulting in the fact the covariance matrix of the measurement noise $\bm R_t$ is an off-diagonal symmetric matrix:
\begin{align}
{\bm R}_t &=
\begin{pmatrix}
{\sigma}_1^2&\cdots&\kappa_{1i}{\sigma}_1{\sigma}_i&\cdots&\kappa_{1m}{\sigma}_1{\sigma}_m\\
\vdots&\ddots&\vdots&\ddots&\vdots\\
\kappa_{1i}{\sigma}_1{\sigma}_i&\cdots&{\sigma}_i^2&\cdots&\kappa_{im}{\sigma}_i{\sigma}_m\\
\vdots&\ddots&\vdots&\ddots&\vdots\\
\kappa_{1m}{\sigma}_1{\sigma}_m&\cdots&\kappa_{im}{\sigma}_i{\sigma}_m&\cdots&{\sigma}_m^2
\end{pmatrix}
\end{align}
where $\kappa_{ij}\in [-1,1]\ (i,j\in\{1,\cdots,m\},\ i\ne j)$ denotes the correlation coefficient for the $i$th and $j$th components of the measurement $\bm y_t$.

In real applications, some components of the correlated $\bm y_t$ may be disturbed by outliers while others are normal. The abnormal value will lead to the performance degradation of the conventional Kalman filters. Directly considering $\bm y_t$ as an outlier and applying the common robust strategies indeed will improve the filtering performance. However, such approaches may result in potential information loss of the normal components. The objective of this work is to propose a unified robust filtering framework for dynamic systems with outlier-contaminated correlated measurements, in which we aim to alleviate the negative effects caused by the abnormal components and meanwhile maximize the utilization of information in normal elements.

\section{Proposed Method}

\subsection{The existing approach}

In the Bayesian filtering paradigm, the posterior density of $\bm x_t$ can be presented by
\begin{align}
p(\bm x_t|\bm y_{1:t})\propto p(\bm x_t|\bm y_{1:t-1})p(\bm y_t|\bm x_t)
\label{bf}
\end{align}
Therefore, under the Gaussian assumption (i.e., both the process and measurement noises are Gaussian) the state estimate at time instant $t$ can be acquired by solving the following optimization problem:
\begin{align}
\hat{\bm x}_{t|t}=\arg\min_{\bm x_t}\left(\mathcal{A}_t+\frac{1}{2}\|\bm y_t-h(\bm x_t)\|^2_{\bm R_t^{-1}}\right)
\label{op_normal}
\end{align}
where $\mathcal{A}_t = \frac{1}{2}\|\bm x_t-\hat{\bm x}_{t|t-1}\|^2_{\bm P_{t|t-1}^{-1}}$ is the model fitting error with the predicted state $\hat{\bm x}_{t|t-1}$ and the associtated prediction error covariance $\bm P_{t|t-1}$. Several Gaussian approximation Kalman filter (GKF) such as the CKF can be applied to efficiently solve the filtering problem in~\eqref{op_normal}.

In some applications, due to unreliable sensors, partial components of correlated measurements may be contaminated by outliers. The existence of these outliers may violate the Gaussian assumption for the measurement noise, resulting in a substantial performance degradation of the conventional Kalman filter. To improve the robustness of filtering algorithms against measurement outliers, the $\ell_2$-norm loss for the MFE in~\eqref{op_normal} is frequently replaced by an outlier-robust cost function, leading to the following robust filtering problem:
\begin{align}
\hat{\bm x}_{t|t}=\arg\min_{\bm x_t}\left(\mathcal{A}_t+\sum_{i=1}^m\rho(\beta_{t,i})\right)
\label{r_prob1}
\end{align}
where $\beta_{t,i}$ is the $i$th component of the NMFE $\bm \beta_t=\bm R_t^{-1/2}(\bm y_t -h(\bm x_t))$, and $\rho(\cdot)$ is a robust cost function such as Huber's cost~\cite{wang2019derivative} and the Hampel cost~\cite{wang2019unified}.

The robust filtering problem in~\eqref{r_prob1} can in principle be solved via a re-weighted approach. Specifically, the optimization problem in~\eqref{r_prob1} can be converted to the following one:
\begin{align}
\hat{\bm x}_{t|t}=\arg\min_{\bm x_t}\left(\mathcal{A}_t+\frac{1}{2}\|\bm y_t-h(\bm x_t)\|^2_{\bm {\bar R}_t^{-1}}\right)
\label{rp2}
\end{align}
with the re-weighted measurement covariance given by
\begin{align}
\bm {\bar R}_t &= \bm R_t^{1/2}\bm W_t^{-1}\bm R_t^{T/2}\label{re}\\
\bm W_t &= \text{diag}\Big(\left[\psi(\beta_{t,1}),\cdots,\psi(\beta_{t,m})\right]\label{re2}\Big)
\end{align}
where $\psi(a)=\rho'(a)/a$ is the associated weight function for the robust cost $\rho(a)$. Some commonly used robust penalty functions and their associated weight function can be found in~\cite{wang2019unified}. Notice that the robust filtering problem~\eqref{r_prob1} shares the similar structure to~\eqref{op_normal} expect for a different weighting matrix, it is therefore~\eqref{r_prob1} can be efficiently solved within the framework of the GKF in an alternating iterative manner.

Although the aforementioned design methodology for robust filtering has been extensively explored, it is not optimal for scenarios where elements of measurements are correlated and meanwhile partial components are contaminated by outliers. A major reason is that the re-weighting operation in~\eqref{re} will affect the variance (or covariance) of the normal components. As a result, some important information about the state may get lost.

This kind of effect reflects in two aspects. For one thing, the NMFE, which is defined by $\bm R_t^{-1/2}(\bm y_t - h(\bm x_t))$, is utilized in the re-weighting operation in~\eqref{re}, and this will enable the outlier-contaminated components to influence the fitting error of normal ones via the off-diagonal matrix $\bm R_t^{-1/2}$. Due to this negative effect, the MFE of the normal components will be misestimated (generally overestimated), leading to an improper reweighting for the variance (or associated covariance) of normal components. For another, the value of the weight function $\psi(\cdot)$ for outlier-contaminated entries is in general relatively small, while that for normal ones is close to $1$. This situation will cause the fact that the reweighted procedure in~\eqref{re} leads to an enlargement of the variance for the normal components.

It is jointly dealing with the MFE of both outlier-contaminated and normal components that results in the misestimation of the variance for the normal measurements. To alleviate this negative effect and mitigate the loss of information, we have proposed a novel reweighting procedure for robust Kalman filtering in which each component of measurements is processed separately in the next section.

\begin{comment}
{\color{red}{The enlargement reflects in two aspects. first the normalized measurement error is calculated by $\bm R_t^{-1/2}(\bm y_t - h(\bm x_t))$, hence the components of the normal index are enlarged since it is effected by outlier components via the off-diagonal matrix $\bm R_t^{-1/2}$. For another, even though the normal and abnormal indices are accurately detected, i.e., $\psi(\mathcal{I})=1$ and $\psi(\mathcal{J})$ is a relative small value, say $10^{-4}$. This reweighted procedure will lead to a enlargement of the variance the normal components. }}

The jointly dealing with the measurement fitting error results in the enlargement phenomenon of the existing method, in the next section we will propose a separated dealing based reweighting procedure for robust Kalman filtering to mitigate the loss of information.
\end{comment}

\subsection{Proposed approach}

Although the measurement $\bm y_t$ conditioned on $\bm x_t$ is correlated via the off-diagonal covariance $\bm R_t$, the marginal distribution of the $i$th component of $\bm y_t$ is given by
\begin{align}
p(y_{t,i}|\bm x_t)\sim\mathcal{N}(h_i(\bm x_t),\sigma_i^2)
\end{align}
where $\sigma_i^2$ is the $i$th diagonal element of $\bm R_t$. Let the MFE be
\begin{align}
\bm \alpha_t = \bm y_t - h(\bm x_t)
\label{imp3}
\end{align}
and the normalized fitting error for $y_{t,i}$ is then defined as
\begin{align}
\delta_{t,i} = \frac{\alpha_{t,i}}{\sigma_i}
\label{asd}
\end{align}
in which $\alpha_{t,i}$ is the $i$th component of $\bm \alpha_t$. If $y_{t,i}$ is disturbed by an outlier, $\delta_{t,i}$ in general tends to be a large value which may degrade the performance of filtering algorithms. Therefore, the variance of $y_{t,i}$ should be reweighed.

Given $\delta_{t,i}$ we can calculate the weighted variance for $i$th measurement component as
\begin{align}
\tilde{\sigma}_i^2 = \frac{\sigma^2_i}{\psi(\delta_{t,i})}
\end{align}
Assuming that the correlation coefficient of any two components remains unchanged, and we can formulate the weighted covariance matrix as:
\begin{align}
\tilde{\bm R}_t\! =\!
\begin{pmatrix}
\tilde{\sigma}_1^2&\cdots&\kappa_{1i}\tilde{\sigma}_1\tilde{\sigma}_i&\cdots&\kappa_{1m}\tilde{\sigma}_1\tilde{\sigma}_m\\
\vdots&\ddots&\vdots&\ddots&\vdots\\
\kappa_{1i}\tilde{\sigma}_1\tilde{\sigma}_i&\cdots&\tilde{\sigma}_i^2&\cdots&\kappa_{im}\tilde{\sigma}_i\tilde{\sigma}_m\\
\vdots&\ddots&\vdots&\ddots&\vdots\\
\kappa_{1m}\tilde{\sigma}_1\tilde{\sigma}_m&\cdots&\kappa_{im}\tilde{\sigma}_i\tilde{\sigma}_m&\cdots&\tilde{\sigma}_m^2
\end{pmatrix}
\label{re_r}
\end{align}
Define
\begin{align}
\bm\Lambda =\text{diag}\Big(\big[\psi(\delta_{t,1})^{-1/2},\cdots,\psi(\delta_{t,m})^{-1/2}\big]\Big)
\label{imp1}
\end{align}
and we can rewrite~\eqref{re_r} into a matrix format as follows:
\begin{align}
\tilde{\bm R}_t=\bm \Lambda\bm R_t\bm \Lambda
\label{imp2}
\end{align}
Replacing $\bar{\bm R}_t$ in~\eqref{rp2} by $\tilde{\bm R}_t$ results in the following novel robust filtering problem
\begin{align}
\hat{\bm x}_{t|t}=\arg\min_{\bm x_t}\left(\mathcal{A}_t+\frac{1}{2}\|\bm y_t-h(\bm x_t)\|^2_{ {\tilde{\bm  R}}_t^{-1}}\right)
\label{r_prob2}
\end{align}

Clearly, the optimization problem~\eqref{r_prob2} has a structure similar to that of the GKF, as presented in~\eqref{op_normal}. However, directly solving~\eqref{r_prob2} within the GKF framework is infeasible due to the fact that $\tilde{\bm  R}_t$ depends on the state $\bm x_t$ via $\delta_{t,i}$ in~\eqref{imp1} and~\eqref{imp2}. To address this, we integrate an alternating iterative procedure into the conventional GKF framework. Specifically, for the given estimate $\hat{\bm x}_{t|t}^k$ after $k$th iteration, we calculate the MFE $\bm \alpha_t$ via~\eqref{imp3}, following the matrices $\bm\Lambda^k$ and $\tilde{\bm R}_t^k$ via~\eqref{imp1} and~\eqref{imp2} respectively. In the next iteration, we solve the optimization problem~\eqref{r_prob2} with $\tilde{\bm R}_t^k$ to obtain $\hat{\bm x}_{t|t}^{k+1}$ by employing the conventional GKF solutions. This iteration loop continues until some stopping criteria are satisfies, e.g.,
\begin{align}
\|\hat{x}_{t \mid t}^{k+1}-\hat{x}_{t \mid t}^{k}\|< \epsilon
\end{align}
for a small tolerance $\epsilon$. At the beginning of the iteration procedure, the predicted state $\bm{\hat x}_{t|t-1}$ is employed to initialize the weighting matrix $\bm\Lambda^0$.

\begin{remark}
The proposed robust filtering framework can be considered as a generalized version of the most commonly utilized robust filtering solutions. Under the same setup (i.e., using the same GKF algorithm and robust cost), the proposed robust filter is equivalent to the existing one when measurements are uncorrelated. Under such a circumstance, we have the entire correlation parameters $\kappa_{ij} = 0 \ (i,j\in \{1,\cdots,m\},\ i\ne j)$, resulting in a diagonal covariance matrix $\bm R_t = \text{diag}([\sigma_1^2,\cdots,\sigma^2_m])$. Therefore, the $i$th component of the NMFE is given by
\begin{align}
\beta_{t,i} = \frac{[\bm y_t-h(\bm x_t)]_i}{\sigma_i}
\end{align}
where $[\bm a]_i$ denotes the $i$th component of a vector $\bm a$. It is apparent that $\beta_{t,i}$ is exactly the same as the one in~\eqref{asd}. In view of the fact that the reweighting procedures of these two robust filters are based on the same value, we can conclude that the proposed robust filter is equivalent to the conventional robust filter, and this conclusion will be verified in the next section.
\end{remark}

\section{Simulation Results}

The proposed algorithm is evaluated by estimating the state of a nonlinear dynamic system. The state-space model of the system is given by
\begin{align}
\bm x_t &=
\begin{pmatrix}
x_{t-1}^{(1)}\text{sin}(x_{t-1}^{(1)})+\text{sin}(x_{t-1}^{(2)})\\
x_{t-1}^{(2)}\text{cos}(x_{t-1}^{(2)})+0.75x_{t-1}^{(1)}
\end{pmatrix}+\bm v_t\\
\bm y_t &=
\begin{pmatrix}
x_t^{(1)}+x_t^{(1)}x_t^{(2)}\\
x_t^{(1)}\text{cos}(2x_t^{(2)})+\text{sin}(x_t^{(1)})
\end{pmatrix} + \bm w_t
\end{align}
where $\bm x_t = [x_t^{(1)},x_t^{(2)}]^T$ is the state of interest, $\bm v_t\sim\mathcal{N}(0,\bm Q_t)$ is the process noise, and $\bm w_t$ is the measurement noise with the nominal covariance matrix $\bm R_t$. The outlier-contaminated measurement noise is generated via the following Gaussian mixture model
\begin{align}
\bm w_t\sim (\bm I -\bm\lambda)\mathcal{N}(0,\bm R_t) + \bm\lambda\mathcal{N}(0,\bm \eta\bm R_t\bm \eta)
\end{align}
where $\bm \lambda = \text{diag}([\lambda_1,\lambda_2])$ is the contamination ratio matrix with $\lambda_i\in[0,1]$ indicating the probability of outlier for $i$th measurement component and $\bm\eta=\text{diag}([\eta_1,\eta_2])$ is a scale factor matrix in which $\eta_i$ is a positive scale to indicate the power of the contaminating noise in the $i$th measurement channel compared with the nominal one.

In this simulation, we consider the Huber cost in our proposed framework. Huber's cost and its associated weight function are given by
\begin{align}
\rho(e)&=\left\{\begin{array}{lc}
0.5e^2,&|e|<\gamma\\
\gamma|e|-0.5\gamma^2,&|e|\ge\gamma
\end{array}\right.\\
\psi(e)&=\left\{\begin{array}{lc}
1,&|e|<\gamma\\
\gamma/|e|,&|e|\ge\gamma
\end{array}\right.
\end{align}
where $\gamma$ is a user-defined parameter which is generally set to $1.345$~\cite{wang2019derivative}. We adopt the CKF as a realization of the GKF, and the resulting method is referred to as the modified Huber-CKF (M-HCKF). We compare the state estimates of the M-HCKF with those of the conventional CKF and the existing Huber CKF (HCKF)~\cite{chang2017unified}. The convergence tolerance $\epsilon$ of the M-HCKF is set to $10^{-6}$. The initial state estimation $\bm {\hat x}_{0|0}$ is generated from $\mathcal{N}(\bm x_0,\bm P_{0})$ with $\bm x_0=[0.5,0.5]^T$ and $\bm P_0=\text{diag}[(0.01,0.01)]$. The nominal covariance matrices of the process and measurement noises are given by
\begin{align}
\bm Q_t = \begin{pmatrix}
0.2&0\\0&0.2
\end{pmatrix},\quad
\bm R_t = \begin{pmatrix}
0.01&\kappa_{12} 0.01\\
\kappa_{12} 0.01&0.01
\end{pmatrix}
\end{align}
where $\kappa_{12}\in[-1,1]$ is the correlation parameter of two measurement components. $\eta_1$ and $\eta_2$ are identically set to $10$. The time-averaged root mean square error (TRMSE) is employed as a performance metric. The TRMSE of $x_t^{(1)}$ is defined as
\begin{align}
\text{TRMSE}_1=\frac{1}{T} \sum_{t=1}^{T} \sqrt{\frac{1}{L} \sum_{j=1}^{L}\left(x_{j, t}^{(1)}-\hat{x}_{j, t}^{(1)}\right)^{2}}
\end{align}
where $j\in\{1,\cdots,L\}$ is the index of the independent Monte Carlo run; $x_{j, t}^{(1)}$ and $\hat{x}_{j, t}^{(1 )}$, respectively, denote the first component of the true and estimated state at time $t$ in the $j$th Monte Carlo run. We do not present the TRMSE of $x_t^{(2)}$ since it has a similar pattern as that of the TRMSE of $x_t^{(1)}$.

\begin{figure}[!h]
\centering
\includegraphics[width=0.34\textwidth]{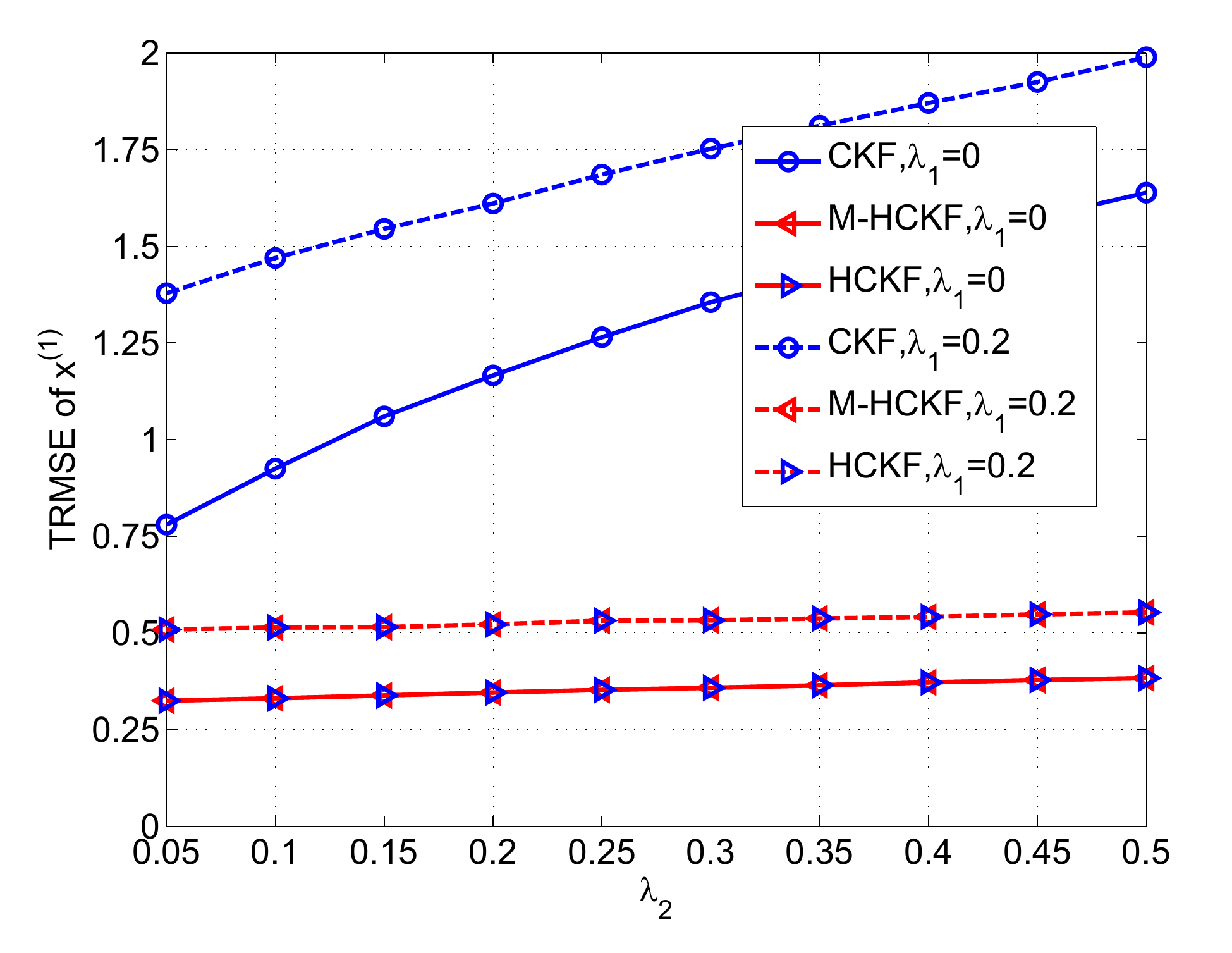}
\caption{TRMSE of $x^{(1)}$ versus $\bm  \lambda$ when $\kappa_{12}=0$.}
\label{f1}
\end{figure}

We first consider the scenario where the correlation parameter $\kappa_{12}=0$ to illustrate the consistency between the M-HCKF and HCKF when measurements are uncorrelated. Here $\lambda_2$ varies from $0.05$ to $0.5$, while $\lambda_1$ is from the set $\{0,0.2\}$. The TRMSE of $x^{(1)}_t$ versus the varied $\bm \lambda$ is presented in Fig.~\ref{f1}, revealing that the robust filters outperform the conventional CKF and meanwhile achieve the same performance.

\begin{figure}[!h]
\centering
\includegraphics[width=0.34\textwidth]{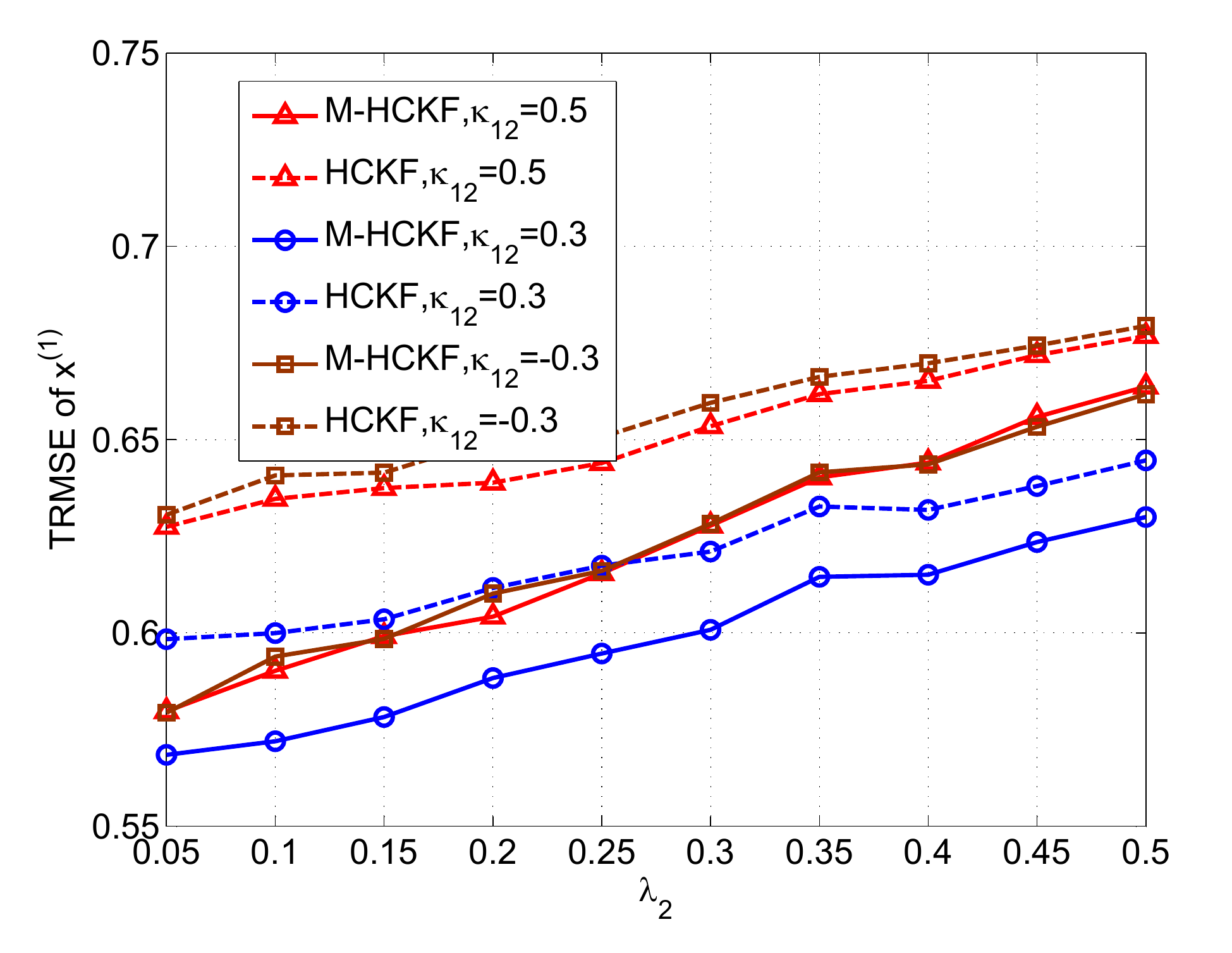}
\caption{TRMSE of $x^{(1)}$ versus $\lambda_2$ with different $\kappa_{12}$ when $\lambda_1=0.2$.}
\label{f2}
\end{figure}

\begin{figure}[!h]
\centering
\includegraphics[width=0.34\textwidth]{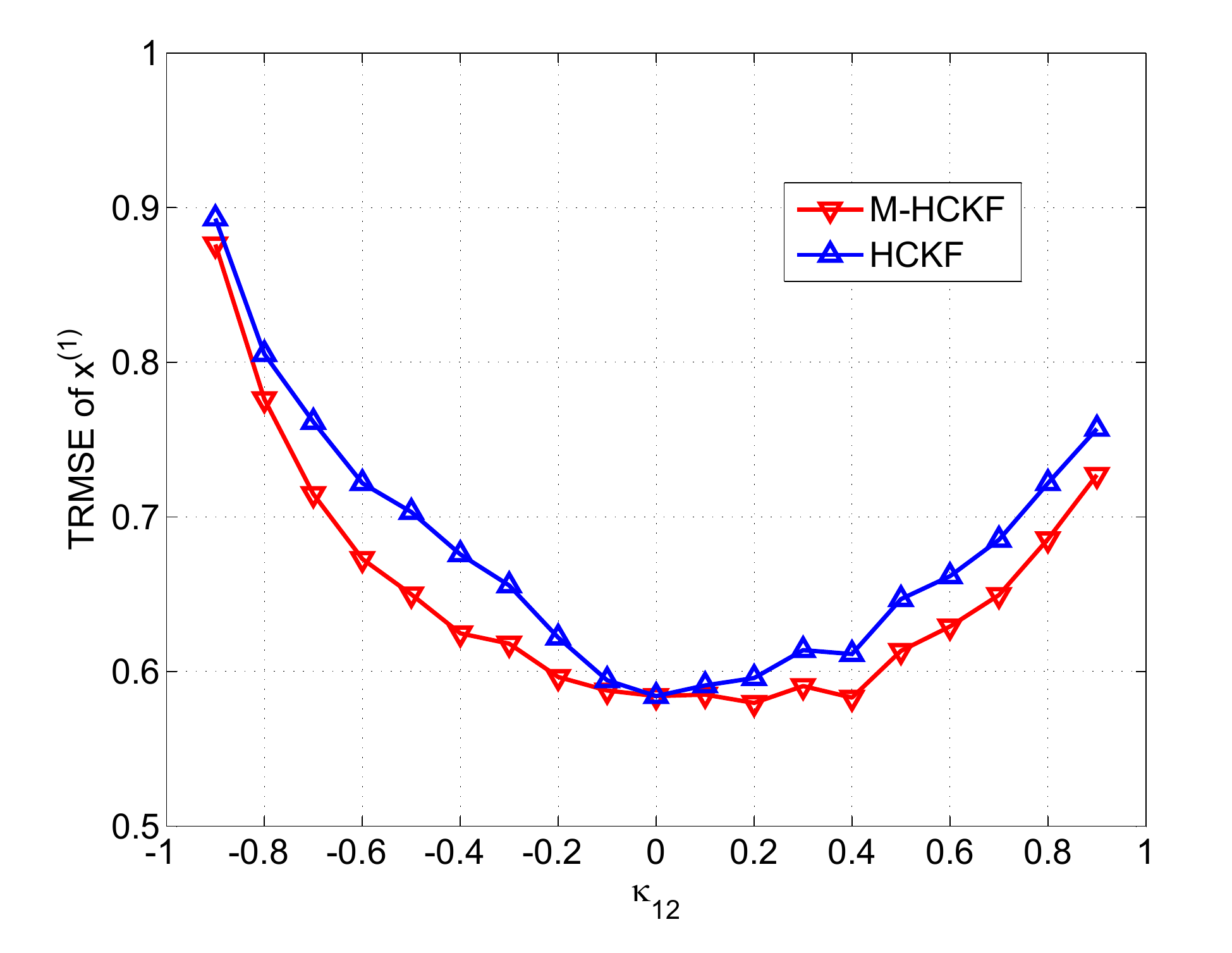}
\caption{TRMSE of $x^{(1)}$ versus $\kappa_{12}$ when $\lambda_1=\lambda_2=0.2$.}
\label{f3}
\end{figure}

Next, we access the performance of these two kinds of robust filters when correlated measurements are involved, and the simulation results are shown in Fig.~\ref{f2} and Fig.~\ref{f3}. We do not include the results of the CKF, since its performance degrades significantly in these scenarios compared with the robust versions. The TRMSE of $x^{(1)}_t$ under different correlation parameters when $\lambda_1=0.2$ and $\lambda_2$ varies is shown in Fig.~\ref{f2}. It is apparent that our proposed method has around $5\%\sim10\%$ performance improvement compared with the HCKF. In Fig.~\ref{f3}, we show the TRMSEs versus the correlation level $\kappa_{12}$ when the outlier contamination ratios $\lambda_1$ and $\lambda_2$ are fixed at $0.2$. It is seen that the M-HCKF has relatively small TRMSEs compared with the HCKF. When $\kappa_{12}= 0$, the TRMSEs of the M-HCKF and HCKF are the same, which is consistent with what we have discussed earlier. Along with the increase of $|\kappa_{12}|$, the gap of the TRMSE between these two robust filters ascends gradually while descends after some specific $\kappa_{12}$.

\section{Conclusions}

We propose a new unified framework for robust Kalman filtering against outliers in the presence of correlated measurements. Unlike the existing robust filtering solutions in which the MFE is jointly processed, we process each component of the MFE separately to avoid the negative effect that the outlier-contaminated elements have enforced on the normal ones. Numerical results show that the proposed method outperforms the existing robust filter when the same GKF and robust cost are utilized. It is noteworthy that although the CKF and Huber's cost are employed in our proposed method in the simulation part, it is straightforward to extend the framework to use with other nonlinear robust penalties (e.g., the Welsch cost~\cite{wang2019unified}) and Gaussian filters (e.g., the UKF~\cite{wan2000unscented}).

%\bibliographystyle{IEEEtran}
%\bibliography{outlier-correlated}

\end{document}